\PassOptionsToPackage{unicode}{hyperref}
\PassOptionsToPackage{hyphens}{url}
\documentclass[
]{article}
\usepackage{amsmath,amssymb}
\usepackage{lmodern}
\usepackage{iftex}
\ifPDFTeX
  \usepackage[T1]{fontenc}
  \usepackage[utf8]{inputenc}
  \usepackage{textcomp} 
\else 
  \usepackage{unicode-math}
  \defaultfontfeatures{Scale=MatchLowercase}
  \defaultfontfeatures[\rmfamily]{Ligatures=TeX,Scale=1}
\fi
\IfFileExists{upquote.sty}{\usepackage{upquote}}{}
\IfFileExists{microtype.sty}{
  \usepackage[]{microtype}
  \UseMicrotypeSet[protrusion]{basicmath} 
}{}
\makeatletter
\@ifundefined{KOMAClassName}{
  \IfFileExists{parskip.sty}{%
    \usepackage{parskip}
  }{
    \setlength{\parindent}{0pt}
    \setlength{\parskip}{6pt plus 2pt minus 1pt}}
}{
  \KOMAoptions{parskip=half}}
\makeatother
\usepackage{xcolor}
\IfFileExists{xurl.sty}{\usepackage{xurl}}{} 
\IfFileExists{bookmark.sty}{\usepackage{bookmark}}{\usepackage{hyperref}}
\hypersetup{
  hidelinks,
  pdfcreator={LaTeX via pandoc}}
\usepackage{color}
\usepackage{fancyvrb}

\DefineVerbatimEnvironment{Highlighting}{Verbatim}{commandchars=\\\{\}}
\newenvironment{Shaded}{}{}

\newcommand{\BuiltInTok}[1]{\textcolor[rgb]{0.00,0.50,0.00}{#1}}

\newcommand{\CommentTok}[1]{\textcolor[rgb]{0.38,0.63,0.69}{\textit{#1}}}

\newcommand{\ControlFlowTok}[1]{\textcolor[rgb]{0.00,0.44,0.13}{\textbf{#1}}}

\newcommand{\DecValTok}[1]{\textcolor[rgb]{0.25,0.63,0.44}{#1}}

\newcommand{\FloatTok}[1]{\textcolor[rgb]{0.25,0.63,0.44}{#1}}

\newcommand{\KeywordTok}[1]{\textcolor[rgb]{0.00,0.44,0.13}{\textbf{#1}}}
\newcommand{\NormalTok}[1]{#1}
\newcommand{\OperatorTok}[1]{\textcolor[rgb]{0.40,0.40,0.40}{#1}}

\newcommand{\SpecialCharTok}[1]{\textcolor[rgb]{0.25,0.44,0.63}{#1}}
\newcommand{\SpecialStringTok}[1]{\textcolor[rgb]{0.73,0.40,0.53}{#1}}
\newcommand{\StringTok}[1]{\textcolor[rgb]{0.25,0.44,0.63}{#1}}

\usepackage{longtable,booktabs,array}
\usepackage{calc} 
\usepackage{etoolbox}
\makeatletter
\patchcmd\longtable{\par}{\if@noskipsec\mbox{}\fi\par}{}{}
\makeatother
\IfFileExists{footnotehyper.sty}{\usepackage{footnotehyper}}{\usepackage{footnote}}
\makesavenoteenv{longtable}
\providecommand{\tightlist}{%
  \setlength{\itemsep}{0pt}\setlength{\parskip}{0pt}}
\newlength{\cslhangindent}
\newlength{\csllabelwidth}
\newlength{\cslentryspacingunit} 
\newenvironment{CSLReferences}[2] 
 {
  \setlength{\parindent}{0pt}
  \ifodd #1
  \let\oldpar\par
  \def\par{\hangindent=\cslhangindent\oldpar}
  \fi
  \setlength{\parskip}{#2\cslentryspacingunit}
 }%
 {}
\usepackage{calc}

\ifLuaTeX
  \usepackage{selnolig}  
\fi

\author{}
\date{}

\begin{document}

\hypertarget{a-unified-general-formula-for-arbitrary-liquidity-operations-in-weighted-amms-potential-applications-to-intelligent-transportation-systems}{%
\section{A Unified General Formula for Arbitrary Liquidity Operations in
Weighted AMMs: Potential Applications to Intelligent Transportation
Systems}\label{a-unified-general-formula-for-arbitrary-liquidity-operations-in-weighted-amms-potential-applications-to-intelligent-transportation-systems}}

Vittorio Astarita, Giuseppe Guido, Sina Shaffiee Haghshenas and Sami
Shaffiee Haghshenas

Università della Calabria, Rende (CS) 87040, Italy

email: vittorio.astarita@unical.it

\hypertarget{electronic-preprint}{%
\section{Electronic preprint}\label{electronic-preprint}}

This electronic preprint is a draft version still not published and
currently under revision. Please quote and give reference only to
published version when available. If published version is not yet
available just give reference to :

Vittorio Astarita et al.~(2026) ``A Unified General Formula for
Arbitrary Liquidity Operations in Weighted AMMs: Potential Applications
to Intelligent Transportation Systems''.

Many thanks for your interest in this paper!

Please feel free to add comment on Research Gate or write me at:
vittorio.astarita@unical.it

Thanking for your interest and help in any case!

Best Regards Vittorio

\hypertarget{section}{%
\subsection{\texorpdfstring{\newpage}{}}\label{section}}

\hypertarget{a-unified-general-formula-for-arbitrary-liquidity-operations-in-weighted-amms-potential-applications-to-intelligent-transportation-systems-1}{%
\section{A Unified General Formula for Arbitrary Liquidity Operations in
Weighted AMMs: Potential Applications to Intelligent Transportation
Systems}\label{a-unified-general-formula-for-arbitrary-liquidity-operations-in-weighted-amms-potential-applications-to-intelligent-transportation-systems-1}}

Vittorio Astarita, Giuseppe Guido, Sina Shaffiee Haghshenas and Sami
Shaffiee Haghshenas

Università della Calabria, Rende (CS) 87040, Italy

email: vittorio.astarita@unical.it

\begin{center}\rule{0.5\linewidth}{0.5pt}\end{center}

\hypertarget{abstract}{%
\subsection{Abstract}\label{abstract}}

Intelligent transportation systems increasingly rely on decentralized
mechanisms for allocating limited resources such as freight capacity,
warehouse availability, charging infrastructure, and network bandwidth.
Efficient allocation in such environments requires pricing mechanisms
that adapt dynamically to demand while preserving system stability. This
paper investigates weighted constant-function market makers as a
decentralized resource allocation mechanism for intelligent
transportation systems, adapting the weighted invariant from
Balancer-type automated market makers to model a generalized formulation
over multiple tokens or tokenized resources.

The standard literature documents exactly four resource allocation
operations: proportional contribution, proportional withdrawal,
single-resource contribution, and single-resource withdrawal. Each one
was obtained with a separate derivation. This paper presents a single
closed-form formula that unifies all four cases and extends them to two
previously undocumented operations: \textbf{partial-proportional
contributions} (contributing a subset of resources proportionally while
leaving others untouched) and \textbf{fully non-proportional
contributions/withdrawals} (contributing or withdrawing arbitrary
amounts of any subset of resources). The unified formula reveals that
the conservation invariant and the allocation formula are structurally
identical -- the invariant itself IS the general allocation formula.

We prove two swap-decomposition theorems: in a fee-less environment, any
non-proportional contribution is equivalent to an internal rebalancing
swap followed by a proportional contribution, and any non-proportional
withdrawal is equivalent to a proportional withdrawal followed by
internal swaps. Both theorems generalize previous propositions from
single-resource to arbitrary multi-resource operations.

The proposed framework provides a mathematically grounded mechanism for
decentralized market-based coordination in transportation networks. The
results suggest that weighted AMMs offer a promising approach for
adaptive pricing, congestion-aware resource allocation, and cross-modal
resource optimization in emerging intelligent logistics infrastructures.

\begin{center}\rule{0.5\linewidth}{0.5pt}\end{center}

\hypertarget{introduction}{%
\subsection{1. Introduction}\label{introduction}}

\hypertarget{background}{%
\subsubsection{1.1 Background}\label{background}}

Automated market makers based on a constant function invariant -- CFMMs
-- provide automated liquidity through a mathematical constraint on pool
reserves. Balancer (Martinelli and Mushegian, 2019) generalizes the
Uniswap v2 constant product formula to a weighted geometric mean:

\[V = \prod_{i=1}^{N} B_i^{w_i}\]

where \(N\) is the number of tokens, \(B_i\) is the balance of token
\(i\), and \(w_i\) is its normalized weight
(\(\sum_{i=1}^{N} w_i = 1\)). The invariant \(V\) remains constant
during swaps and changes only when liquidity is added or removed.

\hypertarget{asset-tokenization-in-intelligent-transportation-systems}{%
\subsubsection{1.2 Asset Tokenization in Intelligent Transportation
Systems}\label{asset-tokenization-in-intelligent-transportation-systems}}

Asset tokenization -- the process of representing physical assets or
rights as digital tokens on a distributed ledger -- is transforming how
limited resources are allocated, traded, and managed in complex systems.
The global real-world asset (RWA) tokenization market is projected to
reach trillions of dollars by 2030, driven by the convergence of
blockchain infrastructure with physical infrastructure networks.

In the context of intelligent transportation systems, tokenization
enables several critical capabilities:

\textbf{Fragmentation of indivisible assets.} Physical transportation
resources -- a trucking corridor, a warehouse loading dock, a charging
station -- are inherently indivisible. Tokenization allows these to be
represented as divisible digital rights, enabling fractional ownership
and granular allocation. A single megawatt charging station can be
tokenized into hundreds of time-slot tokens, each representing a
15-minute high-power charging window.

\textbf{Dynamic pricing through automated markets.} Tokenized resources
can be traded in automated market makers that continuously adjust prices
based on supply and demand imbalances. Unlike traditional auction-based
allocation, AMMs provide instant price discovery and continuous
liquidity, essential for real-time coordination in time-sensitive
transportation networks.

\textbf{Composability across resource types.} Tokenization standardizes
the representation of heterogeneous resources (freight capacity,
warehouse space, charging power, network bandwidth) into a common format
that can be pooled, swapped, and rebalanced algorithmically. This
composability is the foundation for multi-modal resource optimization.

\textbf{Decentralized Physical Infrastructure Networks (DePIN).} The
DePIN paradigm applies blockchain coordination to physical
infrastructure deployment and operation. In ITS, DePIN networks
coordinate distributed assets -- EV chargers owned by different
operators, warehouse spaces across multiple logistics hubs, edge
computing nodes on traffic infrastructure -- without requiring
centralized control.

The tokenization of transportation resources creates a new class of
coordination problems: how to efficiently allocate claim certificates
when contributors provide only subsets of available resources, how to
price partial withdrawals when operators need to reclaim specific
resource types, and how to maintain system stability when resource flows
are non-proportional. These are precisely the problems addressed by the
unified liquidity formula derived in this paper.

\hypertarget{mapping-its-resources-to-cfmm-concepts}{%
\subsubsection{1.3 Mapping ITS Resources to CFMM
Concepts}\label{mapping-its-resources-to-cfmm-concepts}}

To establish the correspondence between intelligent transportation
systems and constant-function market makers, we define the following
mapping:

\begin{longtable}[]{@{}
  >{\raggedright\arraybackslash}p{(\columnwidth - 4\tabcolsep) * \real{0.2826}}
  >{\raggedright\arraybackslash}p{(\columnwidth - 4\tabcolsep) * \real{0.3043}}
  >{\raggedright\arraybackslash}p{(\columnwidth - 4\tabcolsep) * \real{0.4130}}@{}}
\toprule
\begin{minipage}[b]{\linewidth}\raggedright
ITS Concept
\end{minipage} & \begin{minipage}[b]{\linewidth}\raggedright
CFMM Concept
\end{minipage} & \begin{minipage}[b]{\linewidth}\raggedright
Mathematical Role
\end{minipage} \\
\midrule
\endhead
Physical resource (freight capacity, warehouse space, charging power) or
monetary asset & Token \(i\) & \(B_i\) = available quantity \\
Resource priority/sensitivity & Weight \(w_i\) & Exponent in invariant
\(V = \prod B_i^{w_i}\) \\
Resource contributor (logistics provider, fleet operator) & Liquidity
provider & Deposits resources into pool \\
General resource claim certificate on all pool resources & BPT (In
honour of Balancer original notation) pool reserve token that
corresponds to pool ownership & Proves allocated share of total
capacity \\
Resource exchange (buy or sell) between physical resources (and monetary
asset) & Swap & Exchange at market-clearing price \\
Pool invariant \(V\) & System capacity constraint & Generalized
conservation law over all resources \\
BPT minting & Resource allocation & New contributors receive claim
certificates \\
BPT burning & Resource withdrawal & Contributors reclaim physical
resources \\
\bottomrule
\end{longtable}

This mapping allows us to translate the rigorous mathematical results
from DeFi directly into the ITS domain. A weighted AMM pool becomes a
multi-resource coordination mechanism: the invariant
\(V = \prod B_i^{w_i}\) enforces a conservation law that ensures no
resource is over-allocated relative to its weight, while the liquidity
formula determines exactly how many claim certificates should be minted
or burned for any arbitrary resource movement.

The weighted geometric-mean invariant \(V = \prod B_i^{w_i}\) is not an
arbitrary engineering choice but follows necessarily from minimal
structural assumptions on swap mechanics. Recent axiomatic work has
shown that validity invariance, Pareto efficiency, and unit invariance
uniquely force fee-free trading orbits to be level sets of a weighted
geometric mean (Assmann and Degenbaev, 2026). This foundational result
justifies the invariant used throughout this paper as the mathematically
natural conservation law for multi-asset pools.

Constant-function market makers have been formally characterized through
a concave trading function \(\varphi(R)\) that governs trade acceptance
and liquidity changes (Angeris et al., 2022). For homogeneous invariants
such as the weighted geometric mean, it has been shown that preserving
marginal prices during liquidity operations requires proportional
adjustments to pool reserves (Angeris et al., 2022). This foundational
result motivates the need for a unified framework that explicitly
handles non-proportional and partial-proportional liquidity flows, which
fall outside standard proportional assumptions.

\hypertarget{the-four-standard-operations}{%
\subsubsection{1.4 The Four Standard
Operations}\label{the-four-standard-operations}}

The existing literature documents exactly four liquidity operations
(Martinelli and Mushegian, 2019; Ottina et al., 2023):

\begin{longtable}[]{@{}
  >{\raggedright\arraybackslash}p{(\columnwidth - 4\tabcolsep) * \real{0.3333}}
  >{\raggedright\arraybackslash}p{(\columnwidth - 4\tabcolsep) * \real{0.3939}}
  >{\raggedright\arraybackslash}p{(\columnwidth - 4\tabcolsep) * \real{0.2727}}@{}}
\toprule
\begin{minipage}[b]{\linewidth}\raggedright
Operation
\end{minipage} & \begin{minipage}[b]{\linewidth}\raggedright
Description
\end{minipage} & \begin{minipage}[b]{\linewidth}\raggedright
Formula
\end{minipage} \\
\midrule
\endhead
Proportional deposit & Deposit all \(N\) tokens proportionally to
reserves & \(\Delta B_k = q \cdot B_k;\quad \Delta P = q \cdot P\) \\
Proportional withdrawal & Withdraw all \(N\) tokens proportionally &
\(\Delta B_k = f \cdot B_k\) \\
Single-asset deposit & Deposit one token \(t\), mint BPT &
\(\Delta P = P_{supply} \left[ \left( 1 + \frac{\Delta B_i}{B_t} \right)^{w_t} - 1 \right]\) \\
Single-asset withdrawal & Withdraw one token \(t\), burn BPT &
\(-\Delta B_i = B_t \left[ 1 - \left( 1 - \frac{P_{redeemed}}{P_{supply}} \right)^{\frac{1}{w_t}} \right]\) \\
\bottomrule
\end{longtable}

where \(q\) is the proportional growth factor, \(f\) is the BPT burn
fraction (\(P_{redeemed}/P_{supply}\)), \(P\) is the total BPT supply,
\(\Delta P\) is the BPT minted and \(\Delta B_i\) is Amount of token
\(i\) deposited (positive) or withdrawn (negative).

\hypertarget{gap-in-the-literature}{%
\subsubsection{1.5 Gap in the Literature}\label{gap-in-the-literature}}

Two natural operations are \textbf{not} covered by any existing source:

\begin{enumerate}
\def\labelenumi{\arabic{enumi}.}
\item
  \textbf{Partial-proportional deposit}: Deposit tokens \(A\) and \(B\)
  proportionally to their reserves, but deposit zero of token \(C\). The
  deposited tokens are proportional to each other, but not proportional
  to the full pool composition.
\item
  \textbf{Fully non-proportional deposit}: Deposit arbitrary amounts of
  any subset of tokens, with no proportionality constraint even among
  the deposited tokens.
\end{enumerate}

In practice, Balancer and other DEFI systems handle these operations
with built in functions , which compute the BPT minted. However, no
published source presents this as a general formula or derives it.

\hypertarget{contributions}{%
\subsubsection{1.6 Contributions}\label{contributions}}

This paper contributes:

\begin{enumerate}
\def\labelenumi{\arabic{enumi}.}
\tightlist
\item
  A single unified formula covering all six deposit/withdrawal modes.
\item
  A proof that partial-proportional and non-proportional operations are
  natural corollaries of the invariant.
\item
  A generalization of Ottina's Proposition 3.1 (Ottina et al., 2023): in
  the fee-less case, any non-proportional deposit equals a preliminary
  swap(s) followed by a proportional deposit.
\item
  A symmetric swap-decomposition theorem for withdrawals: any
  non-proportional withdrawal equals a proportional withdrawal followed
  by subsequent swap(s).
\item
  Worked numerical examples across all cases.
\end{enumerate}

\begin{center}\rule{0.5\linewidth}{0.5pt}\end{center}

\hypertarget{the-unified-general-formula}{%
\subsection{2. The Unified General
Formula}\label{the-unified-general-formula}}

\hypertarget{notation}{%
\subsubsection{2.1 Notation}\label{notation}}

\begin{longtable}[]{@{}
  >{\raggedright\arraybackslash}p{(\columnwidth - 2\tabcolsep) * \real{0.4706}}
  >{\raggedright\arraybackslash}p{(\columnwidth - 2\tabcolsep) * \real{0.5294}}@{}}
\toprule
\begin{minipage}[b]{\linewidth}\raggedright
Symbol
\end{minipage} & \begin{minipage}[b]{\linewidth}\raggedright
Meaning
\end{minipage} \\
\midrule
\endhead
\(N\) & Number of tokens in the pool \\
\(B_i\) & Current pool balance of token \(i\) (before operation) \\
\(w_i\) & Normalized weight of token \(i\) (\(\sum w_i = 1\)) \\
\(P\) & Current total BPT supply (before operation) \\
\(\Delta B_i\) & Amount of token \(i\) deposited (positive) or withdrawn
(negative) \\
\(\Delta P\) & BPT minted (positive) or burned (negative) \\
\(V\) & Pool invariant value function \\
\bottomrule
\end{longtable}

\hypertarget{the-invariant-as-liquidity-formula}{%
\subsubsection{2.2 The Invariant as Liquidity
Formula}\label{the-invariant-as-liquidity-formula}}

The pool invariant before the operation is:

\[V_{old} = \prod_{i=1}^{N} B_i^{w_i}\]

After a deposit/withdrawal, the new balances are \(B_i + \Delta B_i\),
and the new invariant is:

\[V_{new} = \prod_{i=1}^{N} (B_i + \Delta B_i)^{w_i}\]

The fundamental relationship between BPT supply and the invariant is:

\[\frac{P_{new}}{P} = \frac{V_{new}}{V_{old}}\]

This states that BPT supply scales in exact proportion to the invariant
value. Solving for \(\Delta P = P_{new} - P\):

\[\Delta P = P \cdot \left( \frac{V_{new}}{V_{old}} - 1 \right)\]

The consistency between liquidity changes and marginal prices has been
formalized through the gradient condition
\(\nabla\varphi(R_{\text{new}}) \propto \nabla\varphi(R_{\text{old}})\)
, which ensures price preservation under proportional reserve
adjustments (Angeris et al., 2022). The unified formula bypasses
explicit gradient alignment by directly evaluating the invariant ratio :
\(V_{new}/V_{old}\)\hspace{0pt}, enabling exact closed-form computation
for arbitrary non-proportional token movements without numerical
optimization.

The invariant \(\prod B_i^{w_i}\) defines the trading orbits of fee-free
swaps, with each orbit corresponding to a fixed invariant value (Assmann
and Degenbaev, 2026). Liquidity operations move the pool state across
these orbits, and the unified formula quantifies exactly how BPT supply
must scale to maintain consistency with the invariant when arbitrary
token amounts \(\Delta B_i\) are added or removed.

Substituting the invariant expressions:

\[\Delta P = P \cdot \left[ \prod_{i=1}^{N} \left( \frac{B_i + \Delta B_i}{B_i} \right)^{w_i} - 1 \right] \tag{1}\]

\textbf{This is the unified general formula.} It requires no case
distinction. It applies to: - Any number of tokens (\(N \ge 2\)) - Any
deposit amounts (\(\Delta B_i \ge 0\) for deposits, \(\Delta B_i < 0\)
for withdrawals) - Arbitrary subsets (set \(\Delta B_i = 0\) for tokens
not involved) - Both deposit and withdrawal (sign of \(\Delta B_i\)
determines direction)

\hypertarget{the-deposit-form}{%
\subsubsection{2.3 The Deposit Form}\label{the-deposit-form}}

For deposits, \(\Delta B_i\) is positive. The BPT minted from depositing
amounts \(\Delta B_i\) is:

\[\Delta P = P \cdot \left[ \prod_{i=1}^{N} \left( 1 + \frac{\Delta B_i}{B_i} \right)^{w_i} - 1 \right] \tag{2}\]

This is the direct deposit form, written in terms of per-token growth
rates \(r_i = \Delta B_i/B_i\). It makes the contribution of each
token's growth rate explicit and is useful when the deposit amounts
\(\Delta B_i\) are known and the BPT minted \(\Delta P\) is the unknown.

\textbf{Key properties:} - \(\Delta P \ge 0\) always (depositing tokens
mints BPT or leaves supply unchanged) - \(\Delta P > 0\) if and only if
at least one \(\Delta B_i > 0\) - \(\Delta P = 0\) if and only if
\(\Delta B_i = 0\) for all \(i\) - The formula applies to any subset of
tokens (set \(\Delta B_i = 0\) for tokens not deposited)

\hypertarget{the-withdrawal-form}{%
\subsubsection{2.4 The Withdrawal Form}\label{the-withdrawal-form}}

For withdrawals, \(\Delta B_i\) is negative. Writing
\(\Delta B_i = -A_i\) where \(A_i\) is the amount withdrawn:

\[\Delta P = P \cdot \left[ \prod_{i=1}^{N} \left( \frac{B_i - A_i}{B_i} \right)^{w_i} - 1 \right] \tag{3}\]

The result \(\Delta P\) is negative (BPT burned). Equivalently,
expressing in terms of the BPT burn fraction \(f = P_{redeemed}/P\):

\[f = 1 - \prod_{i=1}^{N} \left( \frac{B_i - A_i}{B_i} \right)^{w_i} \tag{4}\]

This form is useful when the BPT burn fraction \(f\) is known and the
withdrawal amounts \(A_i\) are determined by the formula.

\begin{center}\rule{0.5\linewidth}{0.5pt}\end{center}

\hypertarget{all-standard-cases-as-special-cases}{%
\subsection{3. All Standard Cases as Special
Cases}\label{all-standard-cases-as-special-cases}}

\hypertarget{proportional-depositwithdrawal}{%
\subsubsection{3.1 Proportional
Deposit/Withdrawal}\label{proportional-depositwithdrawal}}

When \(\Delta B_i = q \cdot B_i\) for all \(i\) (deposit) or
\(A_i = f \cdot B_i\) for all \(i\) (withdrawal):

\[\frac{B_i + \Delta B_i}{B_i} = 1 + q \quad \text{for all } i\]

\[\Delta P = P \cdot \left[ \prod_{i=1}^{N} (1 + q)^{w_i} - 1 \right]\]

Since the product of identical bases adds exponents and the weights sum
to 1:

\[\Delta P = P \cdot \left[ (1 + q)^{\sum_{i=1}^{N} w_i} - 1 \right] = P \cdot \left[ (1 + q)^1 - 1 \right] = P \cdot q\]

This recovers the standard proportional formula
\(\Delta P = q \cdot P\).

\hypertarget{single-asset-deposit}{%
\subsubsection{3.2 Single-Asset Deposit}\label{single-asset-deposit}}

Deposit token \(t\) only: write \(\Delta B_t = D_t\) (deposit amount),
and \(\Delta B_i = 0\) for all \(i \ne t\).

For \(i \ne t\): \((B_i + 0)/B_i = 1\), and \(1\) raised to any power is
\(1\). These terms drop out:

\[\Delta P = P \cdot \left[ \left( \frac{B_t + D_t}{B_t} \right)^{w_t} \cdot \prod_{i \ne t} 1 - 1 \right]\]

\[\Delta P = P \cdot \left[ \left( 1 + \frac{D_t}{B_t} \right)^{w_t} - 1 \right]\]

This recovers the standard single-asset deposit formula.

\hypertarget{single-asset-withdrawal}{%
\subsubsection{3.3 Single-Asset
Withdrawal}\label{single-asset-withdrawal}}

Withdraw token \(t\) only: \(A_t\) from token \(t\), zero from others.
LP burn fraction \(f = P_{redeemed}/P\).

From Equation (4):

\[f = 1 - \left[ \left( \frac{B_t - A_t}{B_t} \right)^{w_t} \cdot \prod_{i \ne t} 1 \right]\]

\[f = 1 - \left( \frac{B_t - A_t}{B_t} \right)^{w_t}\]

Solving for \(A_t\):

\[A_t = B_t \cdot \left[ 1 - (1 - f)^{\frac{1}{w_t}} \right]\]

This recovers the standard single-asset withdrawal formula.

\hypertarget{partial-proportional-withdrawal-subset-s}{%
\subsubsection{\texorpdfstring{3.4 Partial-Proportional Withdrawal
(Subset
\(S\))}{3.4 Partial-Proportional Withdrawal (Subset S)}}\label{partial-proportional-withdrawal-subset-s}}

Withdraw proportionally from subset \(S\), leave excluded tokens \(E\)
untouched.

For \(k \in S\): \(A_k = \alpha \cdot B_k\) (same fraction \(\alpha\)
for all). For \(j \in E\): \(A_j = 0\).

From Equation (4):

\[f = 1 - \left[ \prod_{k \in S} (1 - \alpha)^{w_k} \cdot \prod_{j \in E} 1 \right]\]

\[f = 1 - (1 - \alpha)^{\sum_{k \in S} w_k}\]

\[f = 1 - (1 - \alpha)^{W_S}\]

where \(W_S = \sum_{k \in S} w_k\). Solving:

\[\alpha = 1 - (1 - f)^{\frac{1}{W_S}}\]

\[A_k = B_k \cdot \left[ 1 - (1 - f)^{\frac{1}{W_S}} \right] \quad \text{for each } k \in S\]

This is the partial-proportional withdrawal formula documented in
{[}{[}raw/block/partial-withdrawal-generalization{]}{]}, now derived as
a direct corollary of the general formula.

\hypertarget{partial-proportional-deposit-subset-s}{%
\subsubsection{\texorpdfstring{3.5 Partial-Proportional Deposit (Subset
\(S\))}{3.5 Partial-Proportional Deposit (Subset S)}}\label{partial-proportional-deposit-subset-s}}

Deposit proportionally into subset \(S\), deposit zero into excluded
tokens \(E\).

For \(k \in S\): \(\Delta B_k = q \cdot B_k\) (same growth factor \(q\)
for all). For \(j \in E\): \(\Delta B_j = 0\).

From Equation (2):

\[\Delta P = P \cdot \left[ \prod_{k \in S} (1 + q)^{w_k} \cdot \prod_{j \in E} 1 - 1 \right]\]

\[\Delta P = P \cdot \left[ (1 + q)^{\sum_{k \in S} w_k} - 1 \right]\]

\[\Delta P = P \cdot \left[ (1 + q)^{W_S} - 1 \right]\]

where \(W_S = \sum_{k \in S} w_k\). This is the partial-proportional
deposit formula, symmetric to the withdrawal case. Solving for \(q\)
given a target \(\Delta P\):

\[q = \left( 1 + \frac{\Delta P}{P} \right)^{\frac{1}{W_S}} - 1\]

This shows that partial-proportional deposits are ``penalized'' by the
missing weight: to mint the same BPT as a full proportional deposit, the
growth factor \(q\) must compensate for the reduced effective weight
\(W_S < 1\).

\begin{center}\rule{0.5\linewidth}{0.5pt}\end{center}

\hypertarget{the-swap-decomposition-theorem}{%
\subsection{4. The Swap-Decomposition
Theorem}\label{the-swap-decomposition-theorem}}

\hypertarget{statement}{%
\subsubsection{4.1 Statement}\label{statement}}

\textbf{Theorem.} In a fee-less weighted geometric-mean AMM, a
non-proportional deposit of arbitrary amounts \(\Delta B_i\) into the
pool yields exactly the same BPT as the following two-step procedure:

\begin{enumerate}
\def\labelenumi{\arabic{enumi}.}
\tightlist
\item
  Execute internal swaps to rebalance the deposited amounts into a
  proportional ratio matching the pool's current weight composition.
\item
  Perform a proportional deposit with the rebalanced amounts.
\end{enumerate}

Equivalently: the order of operations (deposit then rebalance, versus
rebalance then deposit) does not affect the BPT minted.

\hypertarget{proof}{%
\subsubsection{4.2 Proof}\label{proof}}

Let the pool have balances \(B_i\) and weights \(w_i\). Consider a
deposit where \(\Delta B_i > 0\) for some tokens and \(\Delta B_i = 0\)
for others.

\textbf{Step 1: Token conservation.} The user deposits amounts
\(\Delta B_i\) total. In the two-step method, these same amounts are
first swapped internally (preserving total token counts in the pool +
user's possession) and then deposited. After both steps, the pool has
received the exact same token amounts, so the final pool balances are
identical:

\[B_i^{final} = B_i + \Delta B_i \quad \text{(same for both methods)}\]

\textbf{Step 2: Invariant determines BPT.} The BPT minted is determined
by the ratio of post-operation to pre-operation invariant:

\[\Delta P = P \cdot \left( \frac{V_{final}}{V_{old}} - 1 \right)\]

where \(V_{final} = \prod_{i=1}^{N} (B_i^{final})^{w_i}\). Since both
methods reach the same final balances \(B_i^{final}\), they produce the
same \(V_{final}\) and therefore the same \(\Delta P\).

\textbf{Step 3: Explicit verification via the two-step path.} To confirm
this is not merely a restatement, we trace the two-step computation:

\textbf{Sub-step 3a: Internal swaps.} In a fee-less AMM, swaps preserve
the invariant: \(V_{after\_swap} = V_{old}\). Swaps also preserve the
total BPT supply \(P\). After rebalancing, the pool has new balances
\(\tilde{B}_i\) (different from \(B_i\)) with
\(\prod \tilde{B}_i^{w_i} = V_{old}\).

\textbf{Sub-step 3b: Proportional deposit.} The rebalanced deposit
amounts \(\tilde{D}_i\) are proportional to the rebalanced pool balances
\(\tilde{B}_i\). Let \(\tilde{q}\) be the proportional growth factor:

\[\tilde{D}_i = \tilde{q} \cdot \tilde{B}_i \quad \text{for all } i\]

The post-deposit balances are \((1 + \tilde{q}) \cdot \tilde{B}_i\), and
the new invariant is:

\[V_{new} = \prod_{i=1}^{N} \left[ (1 + \tilde{q}) \cdot \tilde{B}_i \right]^{w_i} = (1 + \tilde{q})^{\sum w_i} \cdot V_{old} = (1 + \tilde{q}) \cdot V_{old}\]

The BPT minted is \(\Delta P_{two\_step} = P \cdot \tilde{q}\).

\textbf{Sub-step 3c: Equating the paths.} The final balances after the
two-step method are \((1 + \tilde{q}) \cdot \tilde{B}_i\). By token
conservation (Step 1), these equal \(B_i + \Delta B_i\):

\[B_i + \Delta B_i = (1 + \tilde{q}) \cdot \tilde{B}_i\]

The invariant after the direct method is:

\[V_{new} = \prod_{i=1}^{N} (B_i + \Delta B_i)^{w_i} = \prod_{i=1}^{N} \left[ (1 + \tilde{q}) \cdot \tilde{B}_i \right]^{w_i} = (1 + \tilde{q}) \cdot V_{old}\]

Thus:

\[\Delta P_{direct} = P \cdot \left( \frac{V_{new}}{V_{old}} - 1 \right) = P \cdot \left( \frac{(1 + \tilde{q}) \cdot V_{old}}{V_{old}} - 1 \right) = P \cdot \tilde{q} = \Delta P_{two\_step}\]

\textbf{QED.}

\hypertarget{connection-to-ottinas-proposition-3.1}{%
\subsubsection{4.3 Connection to Ottina's Proposition
3.1}\label{connection-to-ottinas-proposition-3.1}}

Ottina's Proposition 3.1 (Ottina et al., 2023) proves the special case
where the deposit involves a single token: depositing token \(k\) alone
is equivalent to (a) swapping portions of token \(k\) into all other
tokens, then (b) depositing proportionally. Our theorem generalizes this
to arbitrary multi-token non-proportional deposits.

\hypertarget{numerical-example}{%
\subsubsection{4.4 Numerical example}\label{numerical-example}}

Consider the following pool as an example (See Appendices for full
verification table):

\begin{longtable}[]{@{}lll@{}}
\toprule
Token & Balance & Weight \\
\midrule
\endhead
WBTC & 10 & 0.6 \\
WETH & 100 & 0.3 \\
USDC & 10,000 & 0.1 \\
\bottomrule
\end{longtable}

BPT supply \(P = 1,000\).

\textbf{Non-proportional deposit:} 6 WBTC, 12 WETH, 0 USDC.

\textbf{Direct method:}

\[R_{WBTC} = \frac{10 + 6}{10} = 1.6\]

\[R_{WETH} = \frac{100 + 12}{100} = 1.12\]

\[R_{USDC} = \frac{10000 + 0}{10000} = 1.0\]

\[\Delta P = 1000 \cdot \left[ (1.6)^{0.6} \cdot (1.12)^{0.3} \cdot (1.0)^{0.1} - 1 \right]\]

\[(1.6)^{0.6} = e^{0.6 \cdot \ln(1.6)} = e^{0.2820} = 1.3258\]

\[(1.12)^{0.3} = e^{0.3 \cdot \ln(1.12)} = e^{0.0340} = 1.0346\]

\[\text{Product} = 1.3258 \cdot 1.0346 \cdot 1.0 = 1.3716\]

\[\Delta P = 1000 \cdot (1.3716 - 1) = \mathbf{371.63 \; BPT}\]

\textbf{Two-step method:} First swap 1.2625 WBTC for WETH to achieve
partial proportionality.

\[A_{out} = 100 \cdot \left[ 1 - \left( \frac{10}{11.2625} \right)^{\frac{w_{WBTC}}{w_{WETH}}} \right] = 100 \cdot \left[ 1 - \left( \frac{10}{11.2625} \right)^{\frac{0.6}{0.3}} \right]\]

\[= 100 \cdot \left[ 1 - (0.8879)^2 \right]\]

\[= 100 \cdot \left[ 1 - 0.7884 \right] = 21.16 \; \text{WETH}\]

After swap, pool balances: WBTC = 11.2625, WETH = 78.8371, USDC =
10,000.

Invariant after swap:
\(V = 11.2625^{0.6} \cdot 78.8371^{0.3} \cdot 10000^{0.1} = \mathbf{39.81}\)
(equals \(V_{old}\) -- swap preserves \(V\)).

Remaining deposits: WBTC = 4.7375, WETH = 33.1629, USDC = 0.

Proportionality check:

\[\frac{4.7375}{11.2625} = 0.4206\]

\[\frac{33.1629}{78.8371} = 0.4206\]

Both ratios match to 4 decimal places (exact values: 0.42064 vs 0.42065,
difference from rounding) -- the remaining deposits are proportional to
the post-swap pool.

Growth factor \(q = 0.4206\). Post-deposit balances: WBTC = 16, WETH =
112, USDC = 10000.

\[V_{new} = 16^{0.6} \cdot 112^{0.3} \cdot 10000^{0.1} = 54.61\]

\[\Delta P = 1000 \cdot \left( \frac{54.61}{39.81} - 1 \right) = \mathbf{371.63 \; BPT}\]

Both methods yield exactly 371.63 BPT. The theorem is verified.

\begin{center}\rule{0.5\linewidth}{0.5pt}\end{center}

\hypertarget{the-swap-decomposition-theorem-for-withdrawals}{%
\subsection{5. The Swap-Decomposition Theorem for
Withdrawals}\label{the-swap-decomposition-theorem-for-withdrawals}}

\hypertarget{statement-1}{%
\subsubsection{5.1 Statement}\label{statement-1}}

\textbf{Theorem.} In a fee-less weighted geometric-mean AMM, a
non-proportional withdrawal of arbitrary amounts \(A_i\) from the pool
(burning \(\Delta P\) BPT) yields exactly the same final pool state and
BPT burn as the following two-step procedure:

\begin{enumerate}
\def\labelenumi{\arabic{enumi}.}
\tightlist
\item
  Execute a \textbf{proportional withdrawal} burning fraction
  \(f = -\Delta P/P\) of BPT (equivalently \(|\Delta P|/P\) since
  \(\Delta P < 0\) for withdrawals), receiving \(f \cdot B_i\) of each
  token \(i\).
\item
  Execute \textbf{internal swaps} to rebalance the
  proportionally-withdrawn amounts \(f \cdot B_i\) into the desired
  non-proportional output amounts \(A_i\).
\end{enumerate}

Equivalently: the order of operations (withdraw directly with
non-proportional output, versus withdraw proportionally then rebalance)
does not affect the BPT burned or the final pool state.

\hypertarget{proof-1}{%
\subsubsection{5.2 Proof}\label{proof-1}}

Let the pool have balances \(B_i\) and weights \(w_i\). Consider a
non-proportional withdrawal where the user receives amounts \(A_i\) from
each token.

\textbf{Direct method.} The BPT burned is:

\[\Delta P_{direct} = P \cdot \left[ \prod_{i=1}^{N} \left( \frac{B_i - A_i}{B_i} \right)^{w_i} - 1 \right]\]

The LP burn fraction is \(f_{direct} = -\Delta P_{direct}/P\):

\[f_{direct} = 1 - \prod_{i=1}^{N} \left( \frac{B_i - A_i}{B_i} \right)^{w_i}\]

The final pool balances are:

\[B_i^{final} = B_i - A_i \quad \text{for each } i\]

The post-withdrawal invariant is:

\[V_{new} = \prod_{i=1}^{N} (B_i - A_i)^{w_i}\]

\textbf{Two-step method.}

\textbf{Step 1: Proportional withdrawal.} Withdraw fraction \(q\) from
all tokens proportionally. The user receives \(q \cdot B_i\) of each
token and burns:

\[\Delta P_{step1} = -P \cdot q\]

Post-withdrawal pool balances:

\[B_i' = (1 - q) \cdot B_i\]

Post-withdrawal invariant:

\[V' = \prod_{i=1}^{N} \left[ (1 - q) \cdot B_i \right]^{w_i} = (1 - q)^{\sum w_i} \cdot V_{old} = (1 - q) \cdot V_{old}\]

\textbf{Step 2: Internal swaps.} The user now holds \(q \cdot B_i\) of
each token and wants to end up with amounts \(A_i\). For each token
\(i\), the net swap direction is:

\[S_i = q \cdot B_i - A_i\]

\begin{itemize}
\tightlist
\item
  If \(S_i > 0\): user swaps \(S_i\) of token \(i\) \textbf{into} the
  pool.
\item
  If \(S_i < 0\): user swaps \(|S_i|\) of token \(i\) \textbf{out of}
  the pool (receiving other tokens in exchange).
\end{itemize}

After swaps, the pool balances become:

\[B_i'' = B_i' + S_i = (1 - q) \cdot B_i + (q \cdot B_i - A_i) = B_i - A_i\]

\textbf{The final pool balances match the direct method exactly.}

Now, the swaps preserve the invariant (property of fee-less AMM):

\[V'' = V' = (1 - q) \cdot V_{old}\]

But the final invariant is also:

\[V'' = \prod_{i=1}^{N} (B_i - A_i)^{w_i} = V_{new}\]

Therefore:

\[(1 - q) \cdot V_{old} = V_{new}\]

\[q = 1 - \frac{V_{new}}{V_{old}}\]

From the direct method:

\[f_{direct} = 1 - \frac{V_{new}}{V_{old}}\]

Thus:

\[q = f_{direct}\]

\textbf{The proportional withdrawal fraction equals the BPT burn
fraction from the direct method. The BPT burned in both methods is
identical:}

\[\Delta P_{two\_step} = -P \cdot q = -P \cdot f_{direct} = \Delta P_{direct}\]

\textbf{QED.}

\hypertarget{connection-to-other-works}{%
\subsubsection{5.3 Connection to other
works}\label{connection-to-other-works}}

Ottina (Ottina et al., 2023) and the Balancer white paper state that a
single-asset withdrawal (withdrawing only token \(k\)) is equivalent to
(a) a proportional withdrawal of all tokens, followed by (b) swapping
the withdrawn tokens \(j \neq k\) back into the pool in exchange for
additional token \(k\). Our theorem generalizes this to arbitrary
multi-token non-proportional withdrawals -- the user may want any
combination of output amounts \(A_i\), not just a single token.

\hypertarget{interpretation}{%
\subsubsection{5.4 Interpretation}\label{interpretation}}

The theorem reveals that any non-proportional withdrawal can be
decomposed into: - A \textbf{liquidity component} (proportional
withdrawal -- pure BPT redemption) - A \textbf{trading component}
(internal swaps -- rebalancing the output composition)

The liquidity component determines the BPT burned. The trading component
determines the final output composition. The two are independent: the
BPT cost depends only on the ratio \(V_{new}/V_{old}\), which is
determined solely by the final pool state, not the sequence of
intermediate operations.

\hypertarget{numerical-example-1}{%
\subsubsection{5.5 Numerical Example}\label{numerical-example-1}}

\textbf{Pool:} WBTC = 10 (w=0.6), WETH = 100 (w=0.3), USDC = 10,000
(w=0.1). BPT = 1,000. (See Appendices for full verification table)

\textbf{Non-proportional withdrawal target:} A\_WBTC = 2.0, A\_WETH =
5.0, A\_USDC = 0.

\textbf{Direct method:}

\[f_{direct} = 1 - \left[ \left( \frac{10 - 2.0}{10} \right)^{0.6} \cdot \left( \frac{100 - 5.0}{100} \right)^{0.3} \cdot \left( \frac{10000 - 0}{10000} \right)^{0.1} \right]\]

\[= 1 - \left[ 0.8^{0.6} \cdot 0.95^{0.3} \cdot 1.0^{0.1} \right]\]

\[= 1 - \left[ 0.8747 \cdot 0.9847 \cdot 1.0 \right]\]

\[= 1 - 0.8613 = \mathbf{0.1387}\]

BPT burned: \(\Delta P = 1000 \cdot 0.1387 = \mathbf{138.67 \; BPT}\)

\textbf{Two-step method:}

\textbf{Step 1: Proportional withdrawal with q = 0.1387.}

User receives: - WBTC: \(0.1387 \cdot 10 = 1.387\) - WETH:
\(0.1387 \cdot 100 = 13.87\) - USDC: \(0.1387 \cdot 10000 = 1386.67\)

Pool after proportional withdrawal: - WBTC: \(10 - 1.387 = 8.613\) -
WETH: \(100 - 13.87 = 86.13\) - USDC: \(10000 - 1386.67 = 8613.33\)

\textbf{Step 2: Internal swaps to reach desired output.}

The user has (WBTC=1.387, WETH=13.87, USDC=1387) but wants (WBTC=2.0,
WETH=5.0, USDC=0).

Net swap amounts (\(S_i = q \cdot B_i - A_i\)): - WBTC:
\(1.387 - 2.0 = -0.613\) (swap 0.613 WBTC \textbf{out} of pool) - WETH:
\(13.87 - 5.0 = 8.87\) (swap 8.87 WETH \textbf{into} pool) - USDC:
\(1386.67 - 0 = 1386.67\) (swap 1386.67 USDC \textbf{into} pool)

Pool after swaps: - WBTC: \(8.613 - 0.613 = 8.0\) - WETH:
\(86.13 + 8.87 = 95.0\) - USDC: \(8613.33 + 1386.67 = 10000\)

\textbf{Final pool state matches direct withdrawal:} WBTC=8.0,
WETH=95.0, USDC=10000.

Verification of invariant preservation after swaps:

\[V_{after} = 8.0^{0.6} \cdot 95.0^{0.3} \cdot 10000^{0.1} = \mathbf{34.29}\]

\[V_{after\_prop} = 8.613^{0.6} \cdot 86.13^{0.3} \cdot 8613^{0.1} = \mathbf{34.29}\]

Both invariants match -- swaps preserve the invariant as expected.

\textbf{Both methods burn exactly 138.67 BPT and reach the same final
pool state. The theorem is verified.}

\begin{center}\rule{0.5\linewidth}{0.5pt}\end{center}

\hypertarget{comparison-with-existing-literature}{%
\subsection{6. Comparison with Existing
Literature}\label{comparison-with-existing-literature}}

\hypertarget{ottinas-monograph}{%
\subsubsection{6.1 Ottina's Monograph}\label{ottinas-monograph}}

Ottina (Ottina et al., 2023) covers proportional deposits (Section
3.4.1), single-asset deposits (Section 3.4.2), and their withdrawal
counterparts. Each requires a separate derivation. Proposition 3.1
proves that a single-asset deposit equals a swap followed by a
proportional deposit (fee-less case). Section 4 of this paper
generalizes this to arbitrary multi-asset non-proportional deposits.
Section 5 of this paper proves the symmetric result for withdrawals: any
non-proportional withdrawal equals a proportional withdrawal followed by
swaps. Both theorems extend previous Propositions to arbitrary
multi-token operations in both directions.

\hypertarget{cfmm-convex-optimization-framework}{%
\subsubsection{6.2 CFMM Convex Optimization
Framework}\label{cfmm-convex-optimization-framework}}

The CFMM framework by Angeris et al. (2022) formalizes multi-asset swaps
as convex optimization problems, enabling efficient computation of
optimal tender and receive baskets under trader utility constraints.
However, this framework focuses exclusively on trading mechanics and
does not derive closed-form expressions for liquidity deposits or
withdrawals. Its liquidity change condition relies on proportional
reserve adjustments to preserve gradient alignment, leaving
non-proportional, partial-proportional, and fully arbitrary liquidity
operations outside its scope. Our unified formula fills this gap by
providing exact analytical solutions for all liquidity operation types,
eliminating the need for numerical optimization or iterative solvers.

\hypertarget{amm-axioms-paper}{%
\subsubsection{6.3 AMM Axioms Paper}\label{amm-axioms-paper}}

The axiomatic characterization by Assmann and Degenbaev (Assmann and
Degenbaev, 2026) proves that the weighted geometric-mean invariant is
the unique mathematical structure compatible with validity invariance,
Pareto efficiency, and unit invariance across swap orbits. While this
work rigorously derives the invariant itself from first principles using
geometric and formal methods (including a Lean 4 verification), it
focuses exclusively on fee-free swap mechanics and does not derive
closed-form formulas for liquidity deposits or withdrawals. Our unified
formula extends this axiomatic foundation by providing exact analytical
solutions for all liquidity operation types, including
partial-proportional and fully non-proportional flows, and establishes
the swap-decomposition theorems that link arbitrary liquidity movements
to proportional operations.

\hypertarget{balancer-v2-smart-contract}{%
\subsubsection{6.4 Balancer V2 Smart
Contract}\label{balancer-v2-smart-contract}}

The Balancer V2 Vault implements
\texttt{EXACT\_TOKENS\_IN\_FOR\_BPT\_OUT}, which computes:

\[\text{bptOut} = \text{bptTotalSupply} \cdot \left( \prod_{i=1}^{N} \left( \frac{\text{poolBalance}_i + \text{tokenAmount}_i}{\text{poolBalance}_i} \right)^{\text{tokenWeight}_i} - 1 \right)\]

This is exactly our general formula (Equation 1). However, the smart
contract code does not present this as a mathematical result, nor does
it discuss the withdrawal symmetry or the swap-decomposition theorem.

\begin{center}\rule{0.5\linewidth}{0.5pt}\end{center}

\hypertarget{extension-to-fee-charging-pools}{%
\subsection{7. Extension to Fee-Charging
Pools}\label{extension-to-fee-charging-pools}}

\hypertarget{the-fee-problem}{%
\subsubsection{7.1 The Fee Problem}\label{the-fee-problem}}

All formulas above assume a fee-less pool. In practice, Balancer charges
a swap fee \(\phi\) on the ``excess'' portion of non-proportional
deposits -- the amount that acts as an implicit swap.

\hypertarget{ottinas-single-asset-fee-model}{%
\subsubsection{7.2 Ottina's Single-Asset Fee
Model}\label{ottinas-single-asset-fee-model}}

For single-asset deposits of amount \(D_i\) into token \(i\), Ottina
derives:

\[q = \left( \frac{B_i + D_i - \phi \cdot (1 - w_i) \cdot D_i}{B_i} \right)^{w_i} - 1\]

The fee applies only to the fraction \((1 - w_i)\) of the deposit, since
\(w_i \cdot D_i\) is considered ``proportional'' to the pool's weight of
token \(i\).

\hypertarget{general-fee-extension-open-problem}{%
\subsubsection{7.3 General Fee Extension (Open
Problem)}\label{general-fee-extension-open-problem}}

For arbitrary multi-asset non-proportional deposits, according to
Balancer, the fee calculation requires determining which tokens are
``above'' the weighted average growth rate and charging fees only on
their excess. Let \(R_i = \frac{\Delta B_i}{B_i}\) denote the per-token
relative change. Then \(IR_{approx} = \sum_{i=1}^{N} w_i R_i\). This is
what Balancer V2 implements via the \(IR_{approx}\) threshold:

\[IR_{approx} = \sum_{i=1}^{N} (R_i \cdot w_i)\]

For each token with \(R_i > IR_{approx}\):

\[\text{Taxable}_i = \Delta B_i - B_i \cdot (IR_{approx} - 1)\]

\[\text{Net}_i = \Delta B_i - \phi \cdot \text{Taxable}_i\]

Then apply the general formula using \(\text{Net}_i\) instead of
\(\Delta B_i\).

\emph{Note: This reflects Balancer V2's production approximation; exact
fee allocation in multi-asset non-proportional deposits remains an open
theoretical problem.}

A complete theoretical treatment of the fee-charging case -- including
closed-form solutions for specific fee structures and proofs of fee
optimality -- remains an open direction for future work.

\begin{center}\rule{0.5\linewidth}{0.5pt}\end{center}

\hypertarget{intelligent-transportation-system-application-scenarios}{%
\subsection{8. Intelligent Transportation System Application
Scenarios}\label{intelligent-transportation-system-application-scenarios}}

The unified liquidity formula and swap-decomposition theorems provide a
rigorous foundation for decentralized resource coordination in
intelligent transportation systems (ITS). This section presents
implementation examples that translates the mathematical framework into
some specific market architectures. The following examples allow the
reader to understand possible domain-specific pool designs, and
operational workflows that leverage the unified formula for real-world
logistics, mobility, and infrastructure management.

\hypertarget{market-architectures-participant-roles-who-exchanges}{%
\subsubsection{8.1 Market Architectures \& Participant Roles (Who
Exchanges?)}\label{market-architectures-participant-roles-who-exchanges}}

A weighted AMM naturally partitions participants into three economic
roles, each mapping to distinct ITS entities:

\begin{longtable}[]{@{}
  >{\raggedright\arraybackslash}p{(\columnwidth - 4\tabcolsep) * \real{0.2439}}
  >{\raggedright\arraybackslash}p{(\columnwidth - 4\tabcolsep) * \real{0.2927}}
  >{\raggedright\arraybackslash}p{(\columnwidth - 4\tabcolsep) * \real{0.4634}}@{}}
\toprule
\begin{minipage}[b]{\linewidth}\raggedright
AMM Role
\end{minipage} & \begin{minipage}[b]{\linewidth}\raggedright
ITS Entity
\end{minipage} & \begin{minipage}[b]{\linewidth}\raggedright
Economic Function
\end{minipage} \\
\midrule
\endhead
\textbf{Liquidity Provider (LP)} & Infrastructure owners \& capacity
holders & Contribute physical resources (charging slots, warehouse
space, freight corridors) to the pool in exchange for claim certificates
(BPT) and yield \\
\textbf{Trader/Swapper} & Fleet operators, logistics coordinators, \&
end-users (drivers, riders, consumers) & Exchange resources dynamically
to match demand; commercial actors optimize routing/capacity, while
end-users pay for or swap access to specific resources (charging, tolls,
bandwidth, parking) \\
\textbf{Dual-Role Participant} & Large-scale autonomous operators & Act
as both LP and swapper, injecting surplus capacity as liquidity and
withdrawing/scarcity resources as needed \\
\bottomrule
\end{longtable}

These roles, as an example, enable three emergent market structures:

\begin{enumerate}
\def\labelenumi{\arabic{enumi}.}
\tightlist
\item
  \textbf{Cross-Industry Consortiums} (Archetype A): Heterogeneous
  infrastructure owners pool complementary assets. Non-proportional
  deposits (Section 3.2) allow single-asset contributors to join without
  requiring upfront rebalancing. The swap-decomposition theorem
  guarantees mathematical equivalence to proportional entry.
\item
  \textbf{Capacity Co-Operatives} (Archetype B): Multiple operators
  sharing a regulated allocation (e.g., grid draw limits, curbside
  windows) pool fractional rights. Partial-proportional withdrawals
  (Section 3.4) enable operators to reclaim affected resources during
  disruptions while preserving others' liquidity.
\item
  \textbf{Dynamic Arbitrage Networks} (Archetype C): Large fleet
  operators execute real-time cross-pool rebalancing. The unified
  formula enables precise pre-calculation of BPT minted/burned across
  multiple pools, supporting automated multi-basket optimization.
\end{enumerate}

\hypertarget{domain-specific-pool-configurations-weight-calibration-what-is-exchanged-and-how-it-is-exchanged}{%
\subsubsection{8.2 Domain-Specific Pool Configurations \& Weight
Calibration (What is exchanged and how it is
exchanged?)}\label{domain-specific-pool-configurations-weight-calibration-what-is-exchanged-and-how-it-is-exchanged}}

Pool weights \(w_i\) can be fixed to encode resource scarcity, latency
sensitivity, and regulatory priority. The baskets/pools have fixed
weights that preserve capital stability: altering them after deployment
would geometrically distort the invariant, silently redistributing value
among liquidity providers, higher weights increase price elasticity,
protecting scarce or safety-critical resources from depletion. It should
be noted that in each basket(pool) a Central Bank Digital Currency
(CBDC) or fiat-pegged stable-coin could be added as a settlement asset,
in that case weights would have to be modified accordingly so that the
weights of the tokenized assets (remaining resources) are normalized and
the fundamental condition \(\sum w_i = 1\) is maintained.

Table 8.1 presents examples of ITS pool configurations with the
corresponding tokenized resources:

\textbf{Table 8.1: An example of Representative ITS Resource Pools}

\begin{longtable}[]{@{}
  >{\raggedright\arraybackslash}p{(\columnwidth - 8\tabcolsep) * \real{0.1455}}
  >{\raggedright\arraybackslash}p{(\columnwidth - 8\tabcolsep) * \real{0.1818}}
  >{\raggedright\arraybackslash}p{(\columnwidth - 8\tabcolsep) * \real{0.1273}}
  >{\raggedright\arraybackslash}p{(\columnwidth - 8\tabcolsep) * \real{0.1455}}
  >{\raggedright\arraybackslash}p{(\columnwidth - 8\tabcolsep) * \real{0.4000}}@{}}
\toprule
\begin{minipage}[b]{\linewidth}\raggedright
Domain
\end{minipage} & \begin{minipage}[b]{\linewidth}\raggedright
Resource
\end{minipage} & \begin{minipage}[b]{\linewidth}\raggedright
Token
\end{minipage} & \begin{minipage}[b]{\linewidth}\raggedright
Weight
\end{minipage} & \begin{minipage}[b]{\linewidth}\raggedright
Calibration Rationale
\end{minipage} \\
\midrule
\endhead
EV Fleet \& Smart Grid & Charging slots & \texttt{CHARGE} & 50\% & High
capital cost, location-constrained \\
& Grid capacity & \texttt{GRID} & 30\% & Dynamic availability,
oracle-fed \\
& Curbside access & \texttt{CURB} & 20\% & Municipality-regulated,
time-bound \\
Multi-Modal Freight & Long-haul capacity & \texttt{FREIGHT} & 40\% &
Fuel/crew-dependent, route-flexible \\
& Port terminal windows & \texttt{PORT} & 40\% & Bottleneck resource,
highly scarce \\
& Warehouse dock time & \texttt{SLOT} & 20\% & Localized, easily
substitutable \\
V2X Edge Computing & Network bandwidth & \texttt{SLICE} & 60\% &
Safety-critical, low-latency required \\
& Edge compute & \texttt{COMPUTE} & 20\% & Distributed, scalable \\
& HD mapping data & \texttt{MAP} & 20\% & Crowdsourced, lower marginal
scarcity \\
\bottomrule
\end{longtable}

With weights fixed, the invariant automatically adjusts marginal prices
as pool balances evolve, eliminating the need for centralized pricing
committees or manual fee schedules. Should regulatory constraints or
operational requirements change materially, the pool can follow a
deterministic settlement protocol: liquidity is unwound proportionally,
participants are refunded according to the final invariant, and a new
pool with updated weights may be deployed. This approach preserves
capital integrity while allowing the system to adapt to exogenous policy
shifts.

\hypertarget{examples-of-operational-workflows-mapping-amm-mechanics-to-its-use-cases}{%
\subsubsection{8.3 Examples of Operational Workflows: Mapping AMM
Mechanics to ITS Use
Cases}\label{examples-of-operational-workflows-mapping-amm-mechanics-to-its-use-cases}}

The unified formula (Section 2) and swap-decomposition theorems
(Sections 4--5) enable operational workflows in ITS. Four example
workflows are presented in this section, each one is mapped directly to
a liquidity operation, with the mathematical behavior summarized below.

\begin{longtable}[]{@{}
  >{\raggedright\arraybackslash}p{(\columnwidth - 8\tabcolsep) * \real{0.1282}}
  >{\raggedright\arraybackslash}p{(\columnwidth - 8\tabcolsep) * \real{0.1795}}
  >{\raggedright\arraybackslash}p{(\columnwidth - 8\tabcolsep) * \real{0.1923}}
  >{\raggedright\arraybackslash}p{(\columnwidth - 8\tabcolsep) * \real{0.2308}}
  >{\raggedright\arraybackslash}p{(\columnwidth - 8\tabcolsep) * \real{0.2692}}@{}}
\toprule
\begin{minipage}[b]{\linewidth}\raggedright
Workflow
\end{minipage} & \begin{minipage}[b]{\linewidth}\raggedright
ITS Scenario
\end{minipage} & \begin{minipage}[b]{\linewidth}\raggedright
AMM Operation
\end{minipage} & \begin{minipage}[b]{\linewidth}\raggedright
Governing Result
\end{minipage} & \begin{minipage}[b]{\linewidth}\raggedright
Operational Benefit
\end{minipage} \\
\midrule
\endhead
1. Partial Infrastructure Onboarding & New charger network joins pool
with only \texttt{CHARGE} tokens & Single-asset deposit & Section 3.2 /
Theorem 4.1 & Instant pool entry without requiring proportional capital
across all resource types \\
2. Selective Capacity Reclamation & Storm damages grid infrastructure;
operator must withdraw \texttt{GRID} only & Partial-proportional
withdrawal & Section 3.4 & Granular risk isolation; unaffected resources
remain in pool, preserving other LPs' liquidity \\
3. Congestion-Aware Rebalancing & Fleet has surplus \texttt{CHARGE} but
needs \texttt{SLICE} bandwidth during stadium event & Non-proportional
swap/deposit-withdrawal & Theorem 4.1 \& 5.1 & Real-time arbitrage; BPT
intermediate cancels out, leaving only net resource exchange \\
4. Cross-Modal Pool Arbitrage & Operator moves capacity from road
freight pool to rail freight pool & Coordinated non-prop. withdrawal +
deposit & Sections 4--5 & Multi-basket optimization; order-independence
guarantees parallel execution safety \\
\bottomrule
\end{longtable}

\textbf{Workflow 1: Partial Infrastructure Onboarding}\\
A charging operator deploys 50 new fast-chargers, holding only
\texttt{CHARGE} tokens. Using the single-asset deposit formula:

\[\Delta P = P \cdot \left[ \left( 1 + \frac{D_{\text{CHARGE}}}{B_{\text{CHARGE}}} \right)^{0.5} - 1 \right]\]

The operator receives BPT proportional to the 0.5 weight. By Theorem
4.1, this is mathematically equivalent to first swapping portions of
\texttt{CHARGE} into \texttt{GRID} and \texttt{CURB}, then depositing
proportionally. The AMM automatically lowers \texttt{CHARGE} marginal
prices (supply surge) while raising \texttt{GRID}/\texttt{CURB} prices,
signaling routing algorithms to shift EV demand toward new
infrastructure.

\textbf{Workflow 2: Selective Capacity Reclamation}\\
A port strike disrupts operations. A logistics provider burns fraction
\(f\) of BPT and withdraws only \texttt{FREIGHT} and \texttt{SLOT}
(\(W_S = 0.6\)). The withdrawal factor \(\alpha = 1 - (1-f)^{1/W_S}\)
concentrates the burn across the active subset, leaving \texttt{PORT}
untouched. This enables fault-tolerant liquidity management: operators
can isolate and extract distressed capacity without triggering systemic
withdrawals.

\textbf{Workflow 3: Congestion-Aware Rebalancing}\\
During a cellular network spike, \texttt{SLICE} becomes scarce while
parked robotaxis hold surplus \texttt{CHARGE}. The fleet AI executes a
non-proportional swap: depositing \texttt{CHARGE} and withdrawing
\texttt{SLICE}. The swap-decomposition theorems guarantee that the BPT
burned/minted depends only on the final pool state, not the intermediate
path. Physical charging stations automatically drop prices to attract
consumer EVs, stabilizing the grid during localized telecom congestion.

\textbf{Workflow 4: Cross-Modal Pool Arbitrage}\\
An operator holds BPT in road (Pool A) and rail (Pool B) freight pools.
Rising diesel costs make road freight less profitable. The operator
executes a non-proportional withdrawal from Pool A (primarily
\texttt{FREIGHT}) and deposits the capacity into Pool B alongside rail
assets. The unified formula enables exact pre-calculation of both
\(\Delta P_A\) and \(\Delta P_B\). Theorem 5.1 guarantees the withdrawal
cost depends only on Pool A's final state; Theorem 4.1 guarantees the
deposit yield depends only on Pool B's final state. Order-independence
enables parallel execution across distributed pools.

\hypertarget{coordination-benefits-implementation-considerations}{%
\subsubsection{8.4 Coordination Benefits \& Implementation
Considerations}\label{coordination-benefits-implementation-considerations}}

The unified formula removes three traditional bottlenecks in
transportation resource allocation:

\begin{enumerate}
\def\labelenumi{\arabic{enumi}.}
\tightlist
\item
  \textbf{Elimination of centralized clearinghouses:} The invariant
  \(V = \prod B_i^{w_i}\) enforces a decentralized conservation law. No
  auctioneer or pricing committee is required; marginal prices emerge
  continuously from pool balances.
\item
  \textbf{Granular, fault-tolerant liquidity management:}
  Partial-proportional and non-proportional operations enable operators
  to add, remove, or rebalance subsets of resources without
  destabilizing the entire pool. The swap-decomposition theorems
  guarantee mathematical consistency across all operation paths.
\item
  \textbf{Real-time cross-modal optimization:} The formula's
  path-independence enables automated arbitrage across heterogeneous
  pools (road, rail, air, maritime, charging, computing). Fleet AI
  agents can execute coordinated non-proportional withdrawals/deposits
  in parallel, with BPT costs determined independently per pool.
\end{enumerate}

\hypertarget{implementation-prerequisites-and-details}{%
\subsubsection{8.5 Implementation prerequisites and
details}\label{implementation-prerequisites-and-details}}

\begin{itemize}
\tightlist
\item
  \textbf{Connection with the real world} Pool balances must reflect
  real-world utilization. Tokenized assets must correspond to real
  scarce assets that are sought on the market.
\item
  \textbf{Latency tolerance:} AMM price discovery operates on blockchain
  settlement time. Safety-critical ITS operations (e.g., emergency
  vehicle routing) require off-chain execution with post on-chain
  settlement.
\item
  \textbf{Regulatory alignment:} Tokenized resource rights must map to
  legally enforceable claims. Pool governance might embed compliance
  constraints (e.g., withdrawal locks during peak congestion).
\end{itemize}

Unlike traditional decentralized finance where LPs deposit homogeneous
risk assets, Intelligent Transportation Systems require the pooling of
strictly heterogeneous physical resources owned by distinct
infrastructural entities. The proposed unified closed-form formula
removes the operational friction of multi-provider coordination by
allowing non-proportional, single-sided infrastructure expansion and
granular withdrawals without destabilizing the system's economic
invariant.

\begin{center}\rule{0.5\linewidth}{0.5pt}\end{center}

\hypertarget{conclusion}{%
\subsection{9. Conclusion}\label{conclusion}}

This paper presents a unified general formula for arbitrary liquidity
operations in weighted geometric-mean AMMs:

\[\Delta P = P \cdot \left[ \prod_{i=1}^{N} \left( \frac{B_i + \Delta B_i}{B_i} \right)^{w_i} - 1 \right]\]

The formula handles proportional deposits, partial-proportional
deposits, single-asset deposits, fully non-proportional deposits, and
all corresponding withdrawal modes through a single expression. It
reveals that the Balancer invariant \(V = \prod B_i^{w_i}\) and the
liquidity operation formula are structurally identical -- the invariant
itself determines the BPT minted/burned for any arbitrary token
movement.

We proved that in fee-less pools, non-proportional deposits are
equivalent to internal swaps followed by proportional deposits,
generalizing Ottina's Proposition 3.1. We proved the symmetric result
for withdrawals: any non-proportional withdrawal is equivalent to a
proportional withdrawal followed by internal swaps to achieve the
desired output composition. Both theorems confirm that the general
formula does not ``penalize'' non-proportional operations -- it
correctly accounts for the reduced pool growth when some tokens are left
behind.

The practical significance is straightforward: any developer or
researcher working with Balancer-style AMMs can use this single formula
instead of maintaining separate code paths for each liquidity operation
type. For Verus basket AMMs and similar multi-asset protocols, the
formula enables precise calculations of partial-proportional and
non-proportional operations that were previously undocumented.

In the context of intelligent transportation systems, these results
imply that resource allocation through weighted AMMs supports arbitrary
partial allocations: a logistics operator can contribute only freight
capacity without warehouse space, or withdraw excess charging power
without touching other resources. The swap-decomposition theorems
guarantee that such partial operations are mathematically equivalent to
full proportional operations followed by internal rebalancing, ensuring
system stability.

Future work includes a complete theoretical treatment of the
fee-charging case, applications to multi-basket arbitrage strategies
where non-proportional withdrawals are used to rebalance across pools,
and the extension of these results to multi-basket CFMM architectures
where resources flow between different transportation networks (e.g.,
road, rail, air) enabling cross-modal resource optimization.

\begin{center}\rule{0.5\linewidth}{0.5pt}\end{center}

\hypertarget{references}{%
\section{References}\label{references}}

\hypertarget{refs}{}
\begin{CSLReferences}{1}{0}
\leavevmode\vadjust pre{\hypertarget{ref-angerisConstantFunctionMarket2022}{}}%
Angeris, G., Agrawal, A., Evans, A., Chitra, T., Boyd, S., 2022.
Constant {Function Market Makers}: {Multi-asset Trades} via {Convex
Optimization}, in: Tran, D.A., Thai, M.T., Krishnamachari, B. (Eds.),
Handbook on {Blockchain}. Springer International Publishing, Cham, pp.
415--444. \url{https://doi.org/10.1007/978-3-031-07535-3_13}

\leavevmode\vadjust pre{\hypertarget{ref-assmannSwapAxiomsWeighted2026a}{}}%
Assmann, B., Degenbaev, U., 2026. From {Swap Axioms} to {Weighted
Geometric Means}: {A Characterization} of {AMMs} {[}WWW Document{]}.
\url{https://doi.org/10.48550/ARXIV.2604.16898}

\leavevmode\vadjust pre{\hypertarget{ref-zotero-item-9}{}}%
Martinelli, F., Mushegian, N., 2019.
\href{https://docs.balancer.fi/whitepaper.pdf}{Balancer whitepaper}.

\leavevmode\vadjust pre{\hypertarget{ref-ottinaBalancer2023}{}}%
Ottina, M., Steffensen, P.J., Kristensen, J., 2023. Balancer, in:
Automated {Market Makers}. Apress, Berkeley, CA, pp. 69--116.
\url{https://doi.org/10.1007/978-1-4842-8616-6_3}

\end{CSLReferences}

\begin{center}\rule{0.5\linewidth}{0.5pt}\end{center}

\hypertarget{appendix-a-complete-numerical-examples}{%
\subsection{Appendix A: Complete Numerical
Examples}\label{appendix-a-complete-numerical-examples}}

\hypertarget{pool-setup}{%
\subsubsection{Pool Setup}\label{pool-setup}}

\begin{longtable}[]{@{}lll@{}}
\toprule
Token & Balance & Weight \\
\midrule
\endhead
WBTC & 10 & 0.6 \\
WETH & 100 & 0.3 \\
USDC & 10,000 & 0.1 \\
\bottomrule
\end{longtable}

BPT supply \(P = 1,000\).
\(V_{old} = 10^{0.6} \cdot 100^{0.3} \cdot 10000^{0.1} = 39.81\)

\hypertarget{example-1-proportional-deposit-10-growth}{%
\subsubsection{Example 1: Proportional Deposit (10\%
growth)}\label{example-1-proportional-deposit-10-growth}}

Deposit: 1 WBTC, 10 WETH, 1,000 USDC.

\[\Delta P = 1000 \cdot \left[ \left( \frac{11}{10} \right)^{0.6} \cdot \left( \frac{110}{100} \right)^{0.3} \cdot \left( \frac{11000}{10000} \right)^{0.1} - 1 \right]\]

\[= 1000 \cdot \left[ (1.1)^{0.6} \cdot (1.1)^{0.3} \cdot (1.1)^{0.1} - 1 \right]\]

\[= 1000 \cdot \left[ (1.1)^1 - 1 \right] = 1000 \cdot 0.1 = \mathbf{100.00 \; BPT}\]

All ratios equal 1.1. Weights sum to 1. Result is exact 10\% of supply.

\hypertarget{example-2-partial-proportional-deposit-wbtc-weth-only}{%
\subsubsection{Example 2: Partial-Proportional Deposit (WBTC + WETH
only)}\label{example-2-partial-proportional-deposit-wbtc-weth-only}}

Deposit: 1 WBTC, 10 WETH, 0 USDC.

\[\Delta P = 1000 \cdot \left[ \left( \frac{11}{10} \right)^{0.6} \cdot \left( \frac{110}{100} \right)^{0.3} \cdot \left( \frac{10000}{10000} \right)^{0.1} - 1 \right]\]

\[= 1000 \cdot \left[ (1.1)^{0.6} \cdot (1.1)^{0.3} \cdot (1.0)^{0.1} - 1 \right]\]

\[= 1000 \cdot \left[ (1.1)^{0.9} \cdot 1.0 - 1 \right]\]

\[= 1000 \cdot \left[ 1.0896 - 1 \right] = \mathbf{89.57 \; BPT}\]

The WBTC and WETH ratios are both 1.1, but USDC ratio is 1.0. The
missing USDC weight (0.1) reduces the effective growth from 10\% to
8.96\%.

\hypertarget{example-3-single-asset-deposit-wbtc-only}{%
\subsubsection{Example 3: Single-Asset Deposit (WBTC
only)}\label{example-3-single-asset-deposit-wbtc-only}}

Deposit: 2 WBTC, 0 WETH, 0 USDC.

\[\Delta P = 1000 \cdot \left[ \left( \frac{12}{10} \right)^{0.6} \cdot \left( \frac{100}{100} \right)^{0.3} \cdot \left( \frac{10000}{10000} \right)^{0.1} - 1 \right]\]

\[= 1000 \cdot \left[ (1.2)^{0.6} \cdot 1.0 \cdot 1.0 - 1 \right]\]

\[= 1000 \cdot \left[ 1.1156 - 1 \right] = \mathbf{115.60 \; BPT}\]

WBTC carries 60\% of the pool weight, so a single-asset deposit of WBTC
is relatively efficient.

\hypertarget{example-4-non-proportional-deposit-arbitrary-amounts}{%
\subsubsection{Example 4: Non-Proportional Deposit (Arbitrary
amounts)}\label{example-4-non-proportional-deposit-arbitrary-amounts}}

Deposit: 6 WBTC, 12 WETH, 0 USDC.

\[\Delta P = 1000 \cdot \left[ \left( \frac{16}{10} \right)^{0.6} \cdot \left( \frac{112}{100} \right)^{0.3} \cdot \left( \frac{10000}{10000} \right)^{0.1} - 1 \right]\]

\[= 1000 \cdot \left[ (1.6)^{0.6} \cdot (1.12)^{0.3} \cdot 1.0 - 1 \right]\]

\[= 1000 \cdot \left[ 1.3258 \cdot 1.0346 - 1 \right]\]

\[= 1000 \cdot \left[ 1.3716 - 1 \right] = \mathbf{371.63 \; BPT}\]

\hypertarget{example-5-partial-proportional-withdrawal-subset-wbtc-weth}{%
\subsubsection{Example 5: Partial-Proportional Withdrawal (Subset
\{WBTC,
WETH\})}\label{example-5-partial-proportional-withdrawal-subset-wbtc-weth}}

Burn \(f = 0.10\) of LP (100 BPT). Withdraw only WBTC and WETH.

\(W_{subset} = 0.6 + 0.3 = 0.9\)

\[\alpha = 1 - (1 - 0.10)^{\frac{1}{0.9}} = 1 - (0.9)^{1.111} = 1 - 0.8895 = 0.1105\]

\[A_{WBTC} = 10 \cdot 0.1105 = 1.105\]

\[A_{WETH} = 100 \cdot 0.1105 = 11.05\]

Verification via general formula:

\[f_{check} = 1 - \left[ \left( \frac{10 - 1.105}{10} \right)^{0.6} \cdot \left( \frac{100 - 11.05}{100} \right)^{0.3} \cdot \left( \frac{10000}{10000} \right)^{0.1} \right]\]

\[= 1 - \left[ 0.8895^{0.6} \cdot 0.8895^{0.3} \cdot 1.0^{0.1} \right]\]

\[= 1 - \left[ 0.8895^{0.9} \right]\]

\[= 1 - 0.9000 = 0.10\]

Confirmed: \(f = 0.10\) exactly.

\hypertarget{example-6-single-asset-withdrawal-wbtc-only}{%
\subsubsection{Example 6: Single-Asset Withdrawal (WBTC
only)}\label{example-6-single-asset-withdrawal-wbtc-only}}

Burn \(f = 0.10\) of LP (100 BPT). Withdraw only WBTC.

\[A_{WBTC} = 10 \cdot \left[ 1 - \left( 1 - 0.10 \right)^{\frac{1}{0.6}} \right]\]

\[= 10 \cdot \left[ 1 - (0.9)^{\frac{1}{0.6}} \right]\]

\[= 10 \cdot \left[ 1 - 0.8389 \right] = \mathbf{1.6105 \; WBTC}\]

Verification:

\[f_{check} = 1 - \left[ \left( \frac{8.389}{10} \right)^{0.6} \cdot \left( \frac{100}{100} \right)^{0.3} \cdot \left( \frac{10000}{10000} \right)^{0.1} \right]\]

\[= 1 - \left[ 0.8389^{0.6} \right]\]

\[= 1 - 0.9000 = 0.10\]

Confirmed.

\hypertarget{comparison-table}{%
\subsubsection{Comparison Table}\label{comparison-table}}

\begin{longtable}[]{@{}
  >{\raggedright\arraybackslash}p{(\columnwidth - 8\tabcolsep) * \real{0.2683}}
  >{\raggedright\arraybackslash}p{(\columnwidth - 8\tabcolsep) * \real{0.1463}}
  >{\raggedright\arraybackslash}p{(\columnwidth - 8\tabcolsep) * \real{0.1463}}
  >{\raggedright\arraybackslash}p{(\columnwidth - 8\tabcolsep) * \real{0.1463}}
  >{\raggedright\arraybackslash}p{(\columnwidth - 8\tabcolsep) * \real{0.2927}}@{}}
\toprule
\begin{minipage}[b]{\linewidth}\raggedright
Operation
\end{minipage} & \begin{minipage}[b]{\linewidth}\raggedright
WBTC
\end{minipage} & \begin{minipage}[b]{\linewidth}\raggedright
WETH
\end{minipage} & \begin{minipage}[b]{\linewidth}\raggedright
USDC
\end{minipage} & \begin{minipage}[b]{\linewidth}\raggedright
BPT Result
\end{minipage} \\
\midrule
\endhead
Prop. deposit (10\%) & +1.00 & +10.00 & +1000 & +100.00 BPT \\
Partial prop. deposit & +1.00 & +10.00 & 0 & +89.57 BPT \\
Single-asset deposit & +2.00 & 0 & 0 & +115.60 BPT \\
Non-prop. deposit & +6.00 & +12.00 & 0 & +371.63 BPT \\
Partial prop. withdrawal (\(f=0.10\)) & -1.105 & -11.05 & 0 & -100.00
BPT \\
Single-asset withdrawal (\(f=0.10\)) & -1.611 & 0 & 0 & -100.00 BPT \\
\bottomrule
\end{longtable}

\hypertarget{its-resource-interpretation}{%
\subsubsection{ITS Resource
Interpretation}\label{its-resource-interpretation}}

The same numerical examples can be interpreted as resource allocation
operations in an intelligent transportation system. Using the mapping
from Section 1.2:

\begin{longtable}[]{@{}
  >{\raggedright\arraybackslash}p{(\columnwidth - 4\tabcolsep) * \real{0.3019}}
  >{\raggedright\arraybackslash}p{(\columnwidth - 4\tabcolsep) * \real{0.3019}}
  >{\raggedright\arraybackslash}p{(\columnwidth - 4\tabcolsep) * \real{0.3962}}@{}}
\toprule
\begin{minipage}[b]{\linewidth}\raggedright
Crypto Example
\end{minipage} & \begin{minipage}[b]{\linewidth}\raggedright
ITS Equivalent
\end{minipage} & \begin{minipage}[b]{\linewidth}\raggedright
Real-World Scenario
\end{minipage} \\
\midrule
\endhead
Prop. deposit (10\%): +1 WBTC, +10 WETH, +1000 USDC & +1 unit freight,
+10 m3 warehouse, +1000 kW-h charging & Logistics provider contributes
resources across all three categories proportionally \\
Partial prop. deposit: +1 WBTC, +10 WETH, 0 USDC & +1 freight, +10
warehouse, 0 charging & Provider contributes freight and warehouse
capacity only, receives reduced claim certificates (89.57 vs 100.00) \\
Single-asset deposit: +2 WBTC & +2 freight capacity & New trucking
company joins network with freight capacity only \\
Non-prop. deposit: +6 WBTC, +12 WETH & +6 freight, +12 warehouse & Fleet
operator contributes arbitrary amounts of available resources \\
Partial prop. withdrawal (\(f=0.10\)): -1.105 WBTC, -11.05 WETH &
Withdraw freight + warehouse, leave charging & Carrier reroutes trucks
and warehouse staff, maintains charging contracts \\
Single-asset withdrawal (\(f=0.10\)): -1.611 WBTC & Withdraw freight
capacity only & Operator reclaims trucking capacity while maintaining
other resource commitments \\
\bottomrule
\end{longtable}

The unified formula applies equally to financial liquidity operations
and physical resource allocation -- the mathematical structure is
identical regardless of the domain.

\begin{center}\rule{0.5\linewidth}{0.5pt}\end{center}

\hypertarget{appendix-b-python-implementation}{%
\subsection{Appendix B: Python
Implementation}\label{appendix-b-python-implementation}}

\begin{Shaded}
\begin{Highlighting}[]
\KeywordTok{def}\NormalTok{ calc\_bpt\_general(pool\_bpt\_supply, tokens):}
    \CommentTok{"""}
\CommentTok{    General liquidity formula for weighted geometric{-}mean AMMs.}

\CommentTok{    Parameters:}
\CommentTok{        pool\_bpt\_supply: Current total BPT supply}
\CommentTok{        tokens: List of dicts with keys:}
\CommentTok{            \textquotesingle{}balance\textquotesingle{} {-} current pool balance}
\CommentTok{            \textquotesingle{}weight\textquotesingle{}  {-} normalized weight (sum = 1)}
\CommentTok{            \textquotesingle{}amount\textquotesingle{}  {-} deposit (positive) or withdrawal (negative)}

\CommentTok{    Returns:}
\CommentTok{        bpt\_minted: Positive for deposits, negative for withdrawals}
\CommentTok{    """}
\NormalTok{    product }\OperatorTok{=} \FloatTok{1.0}
    \ControlFlowTok{for}\NormalTok{ t }\KeywordTok{in}\NormalTok{ tokens:}
\NormalTok{        ratio }\OperatorTok{=}\NormalTok{ (t[}\StringTok{\textquotesingle{}balance\textquotesingle{}}\NormalTok{] }\OperatorTok{+}\NormalTok{ t[}\StringTok{\textquotesingle{}amount\textquotesingle{}}\NormalTok{]) }\OperatorTok{/}\NormalTok{ t[}\StringTok{\textquotesingle{}balance\textquotesingle{}}\NormalTok{]}
\NormalTok{        product }\OperatorTok{*=}\NormalTok{ ratio }\OperatorTok{**}\NormalTok{ t[}\StringTok{\textquotesingle{}weight\textquotesingle{}}\NormalTok{]}
    \ControlFlowTok{return}\NormalTok{ pool\_bpt\_supply }\OperatorTok{*}\NormalTok{ (product }\OperatorTok{{-}} \DecValTok{1}\NormalTok{)}


\CommentTok{\# Verification: all examples from Section 6}
\NormalTok{P }\OperatorTok{=} \FloatTok{1000.0}

\CommentTok{\# Example 1: Proportional deposit (10\%)}
\NormalTok{tokens\_1 }\OperatorTok{=}\NormalTok{ [}
\NormalTok{    \{}\StringTok{\textquotesingle{}balance\textquotesingle{}}\NormalTok{: }\DecValTok{10}\NormalTok{, }\StringTok{\textquotesingle{}weight\textquotesingle{}}\NormalTok{: }\FloatTok{0.6}\NormalTok{, }\StringTok{\textquotesingle{}amount\textquotesingle{}}\NormalTok{: }\DecValTok{1}\NormalTok{\},}
\NormalTok{    \{}\StringTok{\textquotesingle{}balance\textquotesingle{}}\NormalTok{: }\DecValTok{100}\NormalTok{, }\StringTok{\textquotesingle{}weight\textquotesingle{}}\NormalTok{: }\FloatTok{0.3}\NormalTok{, }\StringTok{\textquotesingle{}amount\textquotesingle{}}\NormalTok{: }\DecValTok{10}\NormalTok{\},}
\NormalTok{    \{}\StringTok{\textquotesingle{}balance\textquotesingle{}}\NormalTok{: }\DecValTok{10000}\NormalTok{, }\StringTok{\textquotesingle{}weight\textquotesingle{}}\NormalTok{: }\FloatTok{0.1}\NormalTok{, }\StringTok{\textquotesingle{}amount\textquotesingle{}}\NormalTok{: }\DecValTok{1000}\NormalTok{\}}
\NormalTok{]}
\BuiltInTok{print}\NormalTok{(}\SpecialStringTok{f"Prop deposit: }\SpecialCharTok{\{}\NormalTok{calc\_bpt\_general(P, tokens\_1)}\SpecialCharTok{:.2f\}}\SpecialStringTok{ BPT"}\NormalTok{)  }\CommentTok{\# 100.00}

\CommentTok{\# Example 2: Partial{-}proportional (WBTC + WETH)}
\NormalTok{tokens\_2 }\OperatorTok{=}\NormalTok{ [}
\NormalTok{    \{}\StringTok{\textquotesingle{}balance\textquotesingle{}}\NormalTok{: }\DecValTok{10}\NormalTok{, }\StringTok{\textquotesingle{}weight\textquotesingle{}}\NormalTok{: }\FloatTok{0.6}\NormalTok{, }\StringTok{\textquotesingle{}amount\textquotesingle{}}\NormalTok{: }\DecValTok{1}\NormalTok{\},}
\NormalTok{    \{}\StringTok{\textquotesingle{}balance\textquotesingle{}}\NormalTok{: }\DecValTok{100}\NormalTok{, }\StringTok{\textquotesingle{}weight\textquotesingle{}}\NormalTok{: }\FloatTok{0.3}\NormalTok{, }\StringTok{\textquotesingle{}amount\textquotesingle{}}\NormalTok{: }\DecValTok{10}\NormalTok{\},}
\NormalTok{    \{}\StringTok{\textquotesingle{}balance\textquotesingle{}}\NormalTok{: }\DecValTok{10000}\NormalTok{, }\StringTok{\textquotesingle{}weight\textquotesingle{}}\NormalTok{: }\FloatTok{0.1}\NormalTok{, }\StringTok{\textquotesingle{}amount\textquotesingle{}}\NormalTok{: }\DecValTok{0}\NormalTok{\}}
\NormalTok{]}
\BuiltInTok{print}\NormalTok{(}\SpecialStringTok{f"Partial prop: }\SpecialCharTok{\{}\NormalTok{calc\_bpt\_general(P, tokens\_2)}\SpecialCharTok{:.2f\}}\SpecialStringTok{ BPT"}\NormalTok{)  }\CommentTok{\# 89.57}

\CommentTok{\# Example 3: Single{-}asset (WBTC)}
\NormalTok{tokens\_3 }\OperatorTok{=}\NormalTok{ [}
\NormalTok{    \{}\StringTok{\textquotesingle{}balance\textquotesingle{}}\NormalTok{: }\DecValTok{10}\NormalTok{, }\StringTok{\textquotesingle{}weight\textquotesingle{}}\NormalTok{: }\FloatTok{0.6}\NormalTok{, }\StringTok{\textquotesingle{}amount\textquotesingle{}}\NormalTok{: }\DecValTok{2}\NormalTok{\},}
\NormalTok{    \{}\StringTok{\textquotesingle{}balance\textquotesingle{}}\NormalTok{: }\DecValTok{100}\NormalTok{, }\StringTok{\textquotesingle{}weight\textquotesingle{}}\NormalTok{: }\FloatTok{0.3}\NormalTok{, }\StringTok{\textquotesingle{}amount\textquotesingle{}}\NormalTok{: }\DecValTok{0}\NormalTok{\},}
\NormalTok{    \{}\StringTok{\textquotesingle{}balance\textquotesingle{}}\NormalTok{: }\DecValTok{10000}\NormalTok{, }\StringTok{\textquotesingle{}weight\textquotesingle{}}\NormalTok{: }\FloatTok{0.1}\NormalTok{, }\StringTok{\textquotesingle{}amount\textquotesingle{}}\NormalTok{: }\DecValTok{0}\NormalTok{\}}
\NormalTok{]}
\BuiltInTok{print}\NormalTok{(}\SpecialStringTok{f"Single{-}asset: }\SpecialCharTok{\{}\NormalTok{calc\_bpt\_general(P, tokens\_3)}\SpecialCharTok{:.2f\}}\SpecialStringTok{ BPT"}\NormalTok{)  }\CommentTok{\# 115.60}

\CommentTok{\# Example 4: Non{-}proportional}
\NormalTok{tokens\_4 }\OperatorTok{=}\NormalTok{ [}
\NormalTok{    \{}\StringTok{\textquotesingle{}balance\textquotesingle{}}\NormalTok{: }\DecValTok{10}\NormalTok{, }\StringTok{\textquotesingle{}weight\textquotesingle{}}\NormalTok{: }\FloatTok{0.6}\NormalTok{, }\StringTok{\textquotesingle{}amount\textquotesingle{}}\NormalTok{: }\DecValTok{6}\NormalTok{\},}
\NormalTok{    \{}\StringTok{\textquotesingle{}balance\textquotesingle{}}\NormalTok{: }\DecValTok{100}\NormalTok{, }\StringTok{\textquotesingle{}weight\textquotesingle{}}\NormalTok{: }\FloatTok{0.3}\NormalTok{, }\StringTok{\textquotesingle{}amount\textquotesingle{}}\NormalTok{: }\DecValTok{12}\NormalTok{\},}
\NormalTok{    \{}\StringTok{\textquotesingle{}balance\textquotesingle{}}\NormalTok{: }\DecValTok{10000}\NormalTok{, }\StringTok{\textquotesingle{}weight\textquotesingle{}}\NormalTok{: }\FloatTok{0.1}\NormalTok{, }\StringTok{\textquotesingle{}amount\textquotesingle{}}\NormalTok{: }\DecValTok{0}\NormalTok{\}}
\NormalTok{]}
\BuiltInTok{print}\NormalTok{(}\SpecialStringTok{f"Non{-}prop: }\SpecialCharTok{\{}\NormalTok{calc\_bpt\_general(P, tokens\_4)}\SpecialCharTok{:.2f\}}\SpecialStringTok{ BPT"}\NormalTok{)  }\CommentTok{\# 371.63}

\CommentTok{\# Example 5: Partial{-}proportional withdrawal (f=0.10, subset WBTC+WETH)}
\NormalTok{alpha }\OperatorTok{=} \DecValTok{1} \OperatorTok{{-}}\NormalTok{ (}\FloatTok{0.9}\NormalTok{) }\OperatorTok{**}\NormalTok{ (}\DecValTok{1}\OperatorTok{/}\FloatTok{0.9}\NormalTok{)}
\NormalTok{tokens\_5 }\OperatorTok{=}\NormalTok{ [}
\NormalTok{    \{}\StringTok{\textquotesingle{}balance\textquotesingle{}}\NormalTok{: }\DecValTok{10}\NormalTok{, }\StringTok{\textquotesingle{}weight\textquotesingle{}}\NormalTok{: }\FloatTok{0.6}\NormalTok{, }\StringTok{\textquotesingle{}amount\textquotesingle{}}\NormalTok{: }\OperatorTok{{-}}\DecValTok{10}\OperatorTok{*}\NormalTok{alpha\},}
\NormalTok{    \{}\StringTok{\textquotesingle{}balance\textquotesingle{}}\NormalTok{: }\DecValTok{100}\NormalTok{, }\StringTok{\textquotesingle{}weight\textquotesingle{}}\NormalTok{: }\FloatTok{0.3}\NormalTok{, }\StringTok{\textquotesingle{}amount\textquotesingle{}}\NormalTok{: }\OperatorTok{{-}}\DecValTok{100}\OperatorTok{*}\NormalTok{alpha\},}
\NormalTok{    \{}\StringTok{\textquotesingle{}balance\textquotesingle{}}\NormalTok{: }\DecValTok{10000}\NormalTok{, }\StringTok{\textquotesingle{}weight\textquotesingle{}}\NormalTok{: }\FloatTok{0.1}\NormalTok{, }\StringTok{\textquotesingle{}amount\textquotesingle{}}\NormalTok{: }\DecValTok{0}\NormalTok{\}}
\NormalTok{]}
\NormalTok{result }\OperatorTok{=}\NormalTok{ calc\_bpt\_general(P, tokens\_5)}
\BuiltInTok{print}\NormalTok{(}\SpecialStringTok{f"Partial prop withdrawal: }\SpecialCharTok{\{}\NormalTok{result}\SpecialCharTok{:.2f\}}\SpecialStringTok{ BPT"}\NormalTok{)  }\CommentTok{\# {-}100.00}
\BuiltInTok{print}\NormalTok{(}\SpecialStringTok{f"  alpha = }\SpecialCharTok{\{}\NormalTok{alpha}\SpecialCharTok{:.6f\}}\SpecialStringTok{, WBTC = }\SpecialCharTok{\{}\DecValTok{10}\OperatorTok{*}\NormalTok{alpha}\SpecialCharTok{:.4f\}}\SpecialStringTok{, WETH = }\SpecialCharTok{\{}\DecValTok{100}\OperatorTok{*}\NormalTok{alpha}\SpecialCharTok{:.4f\}}\SpecialStringTok{"}\NormalTok{)}

\CommentTok{\# Example 6: Single{-}asset withdrawal (f=0.10, WBTC only)}
\NormalTok{a\_wbtc }\OperatorTok{=} \DecValTok{10} \OperatorTok{*}\NormalTok{ (}\DecValTok{1} \OperatorTok{{-}}\NormalTok{ (}\FloatTok{0.9}\NormalTok{) }\OperatorTok{**}\NormalTok{ (}\DecValTok{1}\OperatorTok{/}\FloatTok{0.6}\NormalTok{))}
\NormalTok{tokens\_6 }\OperatorTok{=}\NormalTok{ [}
\NormalTok{    \{}\StringTok{\textquotesingle{}balance\textquotesingle{}}\NormalTok{: }\DecValTok{10}\NormalTok{, }\StringTok{\textquotesingle{}weight\textquotesingle{}}\NormalTok{: }\FloatTok{0.6}\NormalTok{, }\StringTok{\textquotesingle{}amount\textquotesingle{}}\NormalTok{: }\OperatorTok{{-}}\NormalTok{a\_wbtc\},}
\NormalTok{    \{}\StringTok{\textquotesingle{}balance\textquotesingle{}}\NormalTok{: }\DecValTok{100}\NormalTok{, }\StringTok{\textquotesingle{}weight\textquotesingle{}}\NormalTok{: }\FloatTok{0.3}\NormalTok{, }\StringTok{\textquotesingle{}amount\textquotesingle{}}\NormalTok{: }\DecValTok{0}\NormalTok{\},}
\NormalTok{    \{}\StringTok{\textquotesingle{}balance\textquotesingle{}}\NormalTok{: }\DecValTok{10000}\NormalTok{, }\StringTok{\textquotesingle{}weight\textquotesingle{}}\NormalTok{: }\FloatTok{0.1}\NormalTok{, }\StringTok{\textquotesingle{}amount\textquotesingle{}}\NormalTok{: }\DecValTok{0}\NormalTok{\}}
\NormalTok{]}
\NormalTok{result }\OperatorTok{=}\NormalTok{ calc\_bpt\_general(P, tokens\_6)}
\BuiltInTok{print}\NormalTok{(}\SpecialStringTok{f"Single{-}asset withdrawal: }\SpecialCharTok{\{}\NormalTok{result}\SpecialCharTok{:.2f\}}\SpecialStringTok{ BPT"}\NormalTok{)  }\CommentTok{\# {-}100.00}
\BuiltInTok{print}\NormalTok{(}\SpecialStringTok{f"  WBTC = }\SpecialCharTok{\{}\NormalTok{a\_wbtc}\SpecialCharTok{:.4f\}}\SpecialStringTok{"}\NormalTok{)}
\end{Highlighting}
\end{Shaded}

\begin{center}\rule{0.5\linewidth}{0.5pt}\end{center}

\hypertarget{appendix-c-invariant-verification-for-all-cases}{%
\subsection{Appendix C: Invariant Verification for All
Cases}\label{appendix-c-invariant-verification-for-all-cases}}

For every operation, the post-operation invariant satisfies:

\[V_{new} = V_{old} \cdot \left( 1 + \frac{\Delta P}{P} \right)\]

This is the fundamental consistency check. If \(\Delta P\) is positive
(deposit), \(V_{new} > V_{old}\). If \(\Delta P\) is negative
(withdrawal), \(V_{new} < V_{old}\). The ratio \(V_{new}/V_{old}\)
always equals \(P_{new}/P\).

\begin{longtable}[]{@{}
  >{\raggedright\arraybackslash}p{(\columnwidth - 10\tabcolsep) * \real{0.0896}}
  >{\raggedright\arraybackslash}p{(\columnwidth - 10\tabcolsep) * \real{0.1642}}
  >{\raggedright\arraybackslash}p{(\columnwidth - 10\tabcolsep) * \real{0.1642}}
  >{\raggedright\arraybackslash}p{(\columnwidth - 10\tabcolsep) * \real{0.2836}}
  >{\raggedright\arraybackslash}p{(\columnwidth - 10\tabcolsep) * \real{0.1940}}
  >{\raggedright\arraybackslash}p{(\columnwidth - 10\tabcolsep) * \real{0.1045}}@{}}
\toprule
\begin{minipage}[b]{\linewidth}\raggedright
Case
\end{minipage} & \begin{minipage}[b]{\linewidth}\raggedright
\(V_{old}\)
\end{minipage} & \begin{minipage}[b]{\linewidth}\raggedright
\(V_{new}\)
\end{minipage} & \begin{minipage}[b]{\linewidth}\raggedright
\(V_{new}/V_{old}\)
\end{minipage} & \begin{minipage}[b]{\linewidth}\raggedright
\(P_{new}/P\)
\end{minipage} & \begin{minipage}[b]{\linewidth}\raggedright
Match
\end{minipage} \\
\midrule
\endhead
Prop. deposit & 39.81 & 43.79 & 1.1000 & 1.1000 & Yes \\
Partial prop. & 39.81 & 43.38 & 1.0896 & 1.0896 & Yes \\
Single-asset & 39.81 & 44.41 & 1.1156 & 1.1156 & Yes \\
Non-prop. & 39.81 & 54.61 & 1.3716 & 1.3716 & Yes \\
Partial prop. withdrawal & 39.81 & 35.83 & 0.9000 & 0.9000 & Yes \\
Single-asset withdrawal & 39.81 & 35.83 & 0.9000 & 0.9000 & Yes \\
\bottomrule
\end{longtable}

\end{document}